\documentclass[conference]{IEEEtran}
\IEEEoverridecommandlockouts
\usepackage{cite}
\usepackage{amsmath,amssymb,amsfonts}
\usepackage{algorithmic}
\usepackage{graphicx}
\usepackage{textcomp}
\usepackage{url}
\usepackage{subcaption}
\usepackage{xcolor}
 \usepackage{enumitem}
\def\BibTeX{{\rm B\kern-.05em{\sc i\kern-.025em b}\kern-.08em
    T\kern-.1667em\lower.7ex\hbox{E}\kern-.125emX}}
    
\newcommand{\fig}[1]{Fig.~\ref{fig:#1}}   
  
\newcommand{\tab}[1]{Table~\ref{tab:#1}}  
    
\begin{document}

\title{Microvision: Static analysis-based approach to visualizing microservices in augmented reality 
\thanks{This material is based upon work supported by the National Science Foundation under Grant No. 1854049, grant from Red Hat Research, and Ulla Tuominen (Shapit).}
}

\author{\IEEEauthorblockN{Tomas Cerny}
\IEEEauthorblockA{\textit{\hspace{1em}Department of Computer Science\hspace{1em}} \\
\textit{Baylor University}\\
Waco, Texas, United States \\
tomas\_cerny@baylor.edu}
\and
\IEEEauthorblockN{Amr S. Abdelfattah}
\IEEEauthorblockA{\hspace{1em}\textit{Department of Computer Science\hspace{1em}} \\
\textit{Baylor University}\\
Waco, Texas, United States \\
amr\_elsayed1@baylor.edu}
\and
\IEEEauthorblockN{Vincent Bushong}
\IEEEauthorblockA{\hspace{1em}\textit{Department of Computer Science\hspace{1em}} \\
\textit{Baylor University}\\
Waco, Texas, United States \\
vinbush@gmail.com}
\and
{\,\hspace{3.6em}\,}
\and
\IEEEauthorblockN{Abdullah Al Maruf}
\IEEEauthorblockA{\textit{\hspace{4.6em}Department of Computer Science\hspace{4.6em}} \\
\textit{Baylor University}\\
Waco, Texas, United States \\
maruf\_maruf1@baylor.edu}
\and
\IEEEauthorblockN{Davide Taibi}
\IEEEauthorblockA{\hspace{1.6em}\textit{CloudSEA.AI Group\hspace{1.6em}} \\
\textit{Tampere University}\\
Tampere, FI-33720, Finland \\
davide.taibi@tuni.fi}
\and
{\,\hspace{3.6em}\,}
}

\maketitle

\begin{abstract}
Microservices are supporting digital transformation; however, fundamental tools and system perspectives are missing to better observe, understand, and manage these systems, their properties, and their dependencies. 
Microservices architecture leans toward decentralization, which yields many advantages to system operation; it, however, brings challenges to their development. Microservices lack a system-centric perspective to better cope with system evolution and quality assessment. In this work, we explore microservice-specific architecture reconstruction based on static analysis. Such reconstruction typically results in system models to visualize selected system-centric perspectives. Conventional models are limited in utility when the service cardinality is high. We consider an alternative data visualization using 3D space using augmented reality. To begin testing the feasibility of deriving such perspectives from microservice systems, we developed and implemented prototype tools for software architecture reconstruction and visualization of compared perspectives.
\end{abstract}


\section{Introduction} \label{sec:introduction}


Cloud-native systems frequently use microservices architecture, which revolutionized how we design, develop, and operate software systems. 
The primary goal of microservice architecture is to facilitate the scalability of specific system features, which means dividing the system into self-contained, self-deployable, and easy to scale-out microservices. Best practice guidelines for cloud-native systems, such as Heroku's twelve-factor app\footnote{\url{https://12factor.net}}, suggest practices to build high-quality microservices along with system infrastructure leading to simplified management, monitoring, microservice evolution, resilience, robustness, etc.

With all benefits come drawbacks. Surveys  on microservice practices and challenges~\cite{Bogner2021, SOLDANI2018} revealed common concerns about microservice architecture, including no system-centric view,
problems with overall system evolution, inter-service dependencies, and architectural complexity. 

Each microservice resides in its own self-contained codebase, isolated from the holistic system perspective. This isolation promotes developers' reasoning, they often base the design decisions on the context they are familiar with, which does not necessarily lead to an optimal solution for the overall system. This is where the missing system-centric perspective could help to broaden their reasoning context. 

Recent technologies make it possible to determine system communication paths across microservices using a centralized log or call tracing (e.g., Spring Cloud Sleuth \cite{carnell2021spring}). This could partially determine the system-centric perspective. However, it requires all microservices to be deployed and to operate first. Typically, system endpoints are identified along with inter-dependencies and the dynamic characteristics and metrics of the overall system. However, these technologies are developed for system operation and DevOps engineers. Moreover, they expect user interaction or user simulation tests. Thus, additional efforts are needed to develop tests or wait for user interaction before using these tools. Therefore, such approaches might be challenging to adopt in development pipelines and serve developers, especially considering microservices evolve in an isolated codebase, and we want to reason about the most recent system state, not necessarily the deployed version. 

Moreover, to reveal the white (or at least gray) box system architecture for the system-centric view, we must assess the codebase. 
Suppose we could derive the microservice dependencies solely using static analysis, aggregating results across codebases. As a result, the holistic system reasoning would be greatly simplified and closer to developer practices. The first step to achieve this would require determining the dependencies between microservices~\cite{Cerny2022}.
To address this, we could focus on microservices' bounded contexts \cite{walker2021automatic}, recognize the underlying microservice data models, and then determine data model overlaps based on the similarity of data entities (fields, names, types, etc.) across distinct microservices \cite{closer22}.
In addition, we can identify remote calls with their specific parameters and bind them to other microservice endpoints that match~\cite{das2021automated, closer22}. 

No matter the used approach to reconstruct the software architecture~\cite{cerny2022microservice}, the aggregated information must be presented to practitioners in an understandable way, as the quantity of information can be exhaustive. For example, the Software Architecture Reconstruction process \cite{techRep} recognizes multiple views for different needs; however, the interplay of microservice systems can still be overwhelming. Thus, we must ask a question about the appropriate visualization strategy.

Current approaches that apply dynamic analysis to support the system-centric perspective and seek opportunities in established visual models, typically rendering in the two-dimensional space. We use static analysis to demonstrate it is feasible to derive a system-centric perspective highlighting inter-service dependencies. Moreover, we believe that a more efficient visualization direction should use three-dimensional space to render the system model since it copes better with the ever-growing size requirements of cloud-native systems. We propose static code analysis performed on cloud-native system codebase and its visualization in augmented reality rendered in three-dimensional space. We build on the prototype tool Prophet using static analysis which provides input to the Microvision visualization prototype tool.

The main contribution of this article details how a static analysis-based method for software architecture reconstruction of microservices can be utilized for their visualization. Our approach is with a proof of concept tool tested on a large third-party system testbench. Such reconstruction products can be used for system reasoning. We use it to address the missing system-centric view in these systems that would outline inter-service dependencies. In order to utilize such a reconstruction product, we sought an appropriate presentation suitable for practitioners to easily interpret microservice system details to address common tasks and locate quality aspects. Given the potentially large scale of microservice systems, we research three-dimensional visualization approaches and construct a proof-of-concept visualization in augmented reality, which we compare with conventional two-dimensional representation for the service view. There are benefits and drawbacks to the three-dimensional approach.

The organization of this article is as follows: Section II discusses related work. Section III outlines the Software Architecture Reconstruction (SAR) process. This is followed by Section IV, with conventional and augmented visualization assessed on a third-party large testbench system. 
We assess conventional visualization and augmented visualization in Section V through the use of a small case study. Finally, Section VI draws conclusions and outlines future work.

\section{Background and Related Work} \label{relatedWork}
Software architecture is the central focal point of the system’s development and design. Since systems evolve we must often reconstruct the architecture from the actual system. Software Architecture Reconstruction (SAR) is the process by which the architecture of an implemented system is obtained from the existing system \cite{techRep}. Once we have such architecture reconstructed we can reason about the system (i.e, conformance checking, verification, evolution, modification, extensions, documentation generation, etc.), or visualize the system. Since software architecture is complex and interpreted differently for various system aspects, it can be described by architectural views \cite{10.1145/141874.141884, Bass2012}. These views capture certain system qualities or aspects. The foundation for successful SAR is the ability to reconstruct effective architectural views of a system \cite{walker2021automatic}. Existing SAR work related to microservices by Rademacher et al. \cite{10.1007/978-3-030-49418-6_21} has considered four views as their outcome. In particular, it operated with domain, technology, service, and operation views. Each of these views considered a specific perspective and related concerns within the system. However, each also relates to other views. As an example, consider the service view overlapping with the domain view to detail which data entities are involved in endpoints. The technology and domain view will then show where the data entities persist. 

\subsection{Static and Dynamic Analysis Visualization}
The process of Software Architecture Reconstruction can be exhaustive with manual efforts or involve automation \cite{10.1007/978-3-030-49418-6_21}. Given that microservices are decentralized, the whole process becomes even more complicated since multiple codebases can be involved. Rademacher et al. \cite{10.1007/978-3-030-49418-6_21} manually collected architecture-related artifacts, constructed a canonical representation of the data model, and based on that fused module views. They then performed architecture analysis on the results to answer hypotheses about architecture implementation from the reconstructed architecture information. 

Considering current practices in microservice development and operations (DevOps), {\em dynamic} analysis can be used involving tracing. Tracing adds a tracing identifier to log messages generated throughout the system interaction, which allows us to centralize these messages via centralized logging and interpret their content in the holistic context by observing inter-microservice dependencies. It is common to extract dependency graphs such as directed acyclic graphs~\cite{al2022using}. The advantage of this approach is platform independence.  However, received log messages and their origins only lead to an abstract reconstruction providing more or less a black-box view. The industry practice provides monitoring, tracing, and metrics tools to capture data about the microservices \cite{carnell2021spring} (i.e., OpenTelemetry, Zipkin, or Jaeger \footnote{https://opentelemetry.io; https://zipkin.io; http://jaegertracing.io}).
These tools seamlessly integrate with enterprise frameworks and utilize existing mechanisms such as method call interception or instrumentation. Mayer and Weinreich use the Spring framework's interceptors to monitor runtime calls between services to generate an architectural view of a microservice system \cite{interceptor_extraction}. However, it is also possible to use API gateway \cite{10.1145/3147213.3147229}. 
Granatelli et al. \cite{recovering_architecture} approached the challenge by querying the containerization framework to retrieve calls between microservices at runtime.

The {\em static} analysis perspective can be constructed from artifacts available before deployment. Analyzing a program's codebases has played a part in the formal verification of a system's correctness \cite{Chlipala:2013:BSP:2544174.2500592, 10.1007/978-3-540-69611-7_8} and other fields. However, the major challenge is the decentralized codebase. In the realm of microservices, it has been used to identify calls between microservices to generate security policy automatically \cite{auto_authorization}. Also, it has been used to analyze monolithic applications to recommend splits for converting to microservices \cite{auto_extract_graphs,Taibi2019closer19,Azadi2019}. 
In generating a service dependency graph, Esparrachiari et al. \cite{Esparrachiari:2018:TCM:3277539.3277541} posit that source code analysis is not sufficient since the deployment environment may impact the actual dependencies a given deployed module has. Pigazzini et al. \cite{Pigazzini2020} reconstructed the architecture of microservices-based systems parsing Java source files and Docker/Spring configuration files, with the goal of identifying cyclic dependencies between microservices. However, related works mainly focused on the identification of the anti-patterns\cite{Taibi2018IEEE, Taibi2020MSE} proposing a visualization for the system architecture. Rahman et al. \cite{Rahman2019} followed a similar approach to parse the code, nevertheless, they developed a tool named "MicroDepGraph"\footnote{MicroDepGraph \url{https://github.com/clowee/MicroDepGraph}} to visualize the call-graph between microservices. Ibrahim et al. used a project's Dockerfiles to search for known security vulnerabilities of the container images being used, which they overlay on the system topology extracted from Docker Compose files to generate an attack graph showing how a security breach could be propagated through a microservice mesh \cite{attack_graph}.
In preliminary work \cite{closer22} we used the approach proposed by Rademacher et al. \cite{10.1007/978-3-030-49418-6_21} and demonstrated that automated merge of microservices data models or detection of inter-service communication is feasible.
In our recent work, we highlighted the power of static and dynamic analysis for detecting microservices API Patterns~\cite{bakhtin2022survey}. 

\subsection{Architecture Visualization}

Zhou et al. \cite{9097240} sought common visualizations for enterprise architectures. The most common directions are ArchiMate, UML, Business Motivation Model (BMM), and BPMN, among others. The Open Group Architecture Framework (TOGAF) is the most frequently used framework for enterprise architecture, further extended by the Architectural Development Method using ArchiMate. It is typically modeled at four levels with different specializations: Business, Application, Data, and Technology, which to some extent correspond to the architectural views described by Rademacher et al. \cite{10.1007/978-3-030-49418-6_21}, with the exception of business architecture levels which rather drives the motivation for the implementation.

The C4 model (Context, Containers, Components, and Code) is a practical approach for modeling software architecture \cite{c4} given a hierarchical model consisting of four levels of abstraction, ranging from the high-level system context to individual code elements. Alternative visualization practices have emerged for software architectures \cite{Shahin2014ASR, 10.1145/1409720.1409745}. Shahin et al. \cite{Shahin2014ASR} categorize alternative visualization as graph-based visualization involving graphs showing nodes and links similar to ontologies. Another approach is a notation-based visualization, such as UML or SysML, or Matrix-based approaches. Quite common is a metaphor-based visualization that uses familiar physical world contexts (e.g., cities, islands, or landscapes). To make the system more understandable using a visual metaphor, Virtual and Augmented Reality (VR/AR) methods have been explored for software architecture visualization. One example is a "software city"; software packages are represented as buildings and their dependencies as streets, which is an example of virtual reality \cite{vr_software_cities, vr_evostreets,vr_distributed}. Schreiber et al. proposed to show individual software modules as "islands" in an ocean displayed in AR. Software packages and classes in each module are represented as regions and buildings on the module island, and, importantly, module imports and exports are displayed as ports that connect the different islands. The VR-EA tool from Oberhauser et al. \cite{vrea} is an attempt to visualize larger enterprise applications. However, it uses modeling tools as inputs to generate a 3D VR view in the virtual reality of business processes and their relationships with enterprise resources. While it can show a large group of interconnected components, it depends on a set of models that must be manually created. 


\begin{figure*}[t!]
\centering
\includegraphics[width=0.85\textwidth]{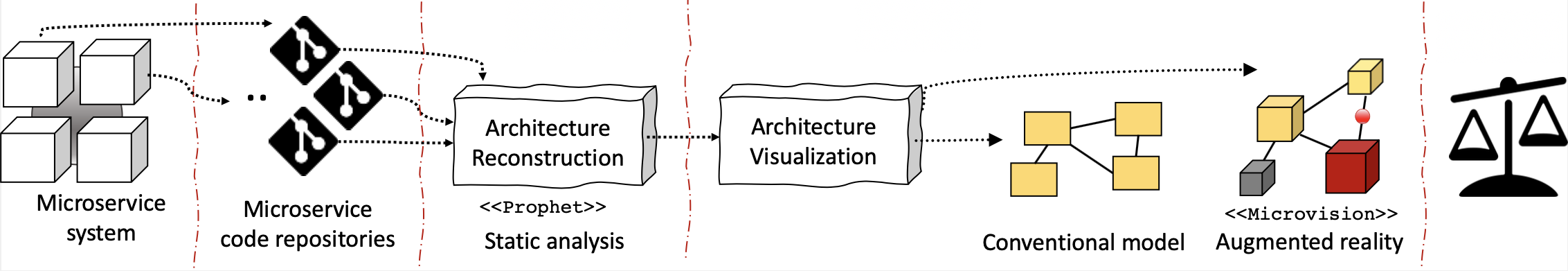}
\vspace{-0.5em}
\caption{Microvision Construction Process}
\label{fig:process}
\vspace{-1.5em}
\end{figure*}

\section{Static analysis-based SAR of Microservices} \label{sec:sar_process}

Microservice systems are by nature decentralized. The Heroku's 12-factor app methodology \cite{TheTwelv82:online,carnell2021spring,Pigazzini2020} recommends that each microservice be 
self-contained with its own codebase and database to facilitate and improve evolution, scalability, and dependency management. However, microservices are not isolated; they interact using interfaces or message queues. Thus, there is a dependency between microservices; however, it is typically a loose one. Since the interaction happens through interfaces, perhaps the most notable dependency is on the endpoint names and the parameters that represent data or transfer objects. Based on domain-driven development \cite{10.1007/978-3-030-49418-6_21,Cerny:2018:CUM:3183628.3183631}, each module considers a bounded context \cite{carnell2021spring}, which includes a limited perspective of the system holistic data model called context map, and often bounded contexts partially
\begin{figure}[b!]
\vspace{-1em}
\centering\includegraphics[width=22.8em]{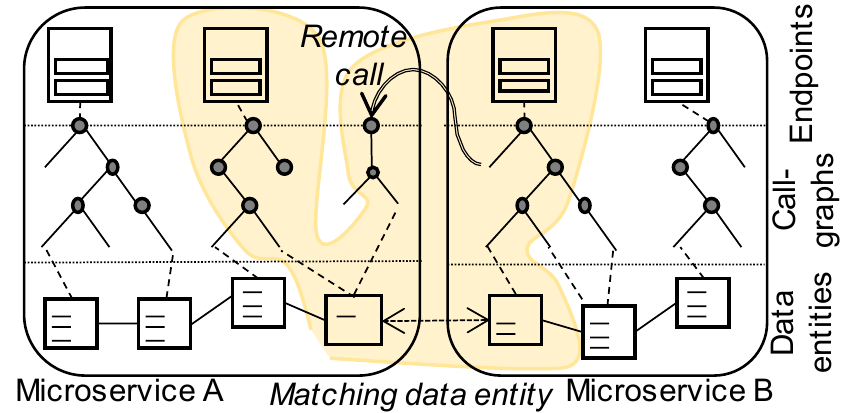}
\vspace{-.51em}
\caption{Microservice dependencies}
\label{fig:merge}
\vspace{-.5em}
\end{figure}
overlap through certain data entities with other modules. We can use this overlap as an ingredient 
to determine the system-centric view. Apart from this, the inter-service interaction, such as REST/RPC endpoint calls, is 
another ingredient we can consider. These two strategies are illustrated in \fig{merge}. 

Our SAR process considers the static-code analysis of individual microservice's codebases. As suggested by Carnell et al. \cite{carnell2021spring}, the codebase contains the source code and build and deployment configuration files possibly relevant to the process. 

Our process is illustrated in \fig{process}, it starts with the \textit{extraction} phase, operating with Abstract Syntax Trees (AST) parsed from the code. Next, walking through the tree and recognizing method calls we 
extract call-graphs. Top methods are candidates for endpoints; in addition, frameworks typically augment these endpoints with additional information (i.e., HTTP types, constraints, etc.) that indicate the endpoint (i.e., in the form of annotations or external files). With endpoints detected, we assess their parameters and trace the calls down through the controllers, services, and repositories to referenced and involved data entities. We detect these components' types by assessing the associated properties in the AST and the call-graph from the endpoint. This tracing also allows us to derive dependencies between endpoints and involved data entities. We determine which endpoints operate with specific data entities in the reverse perspective. Walking through the call-graphs, we can detect involved constraints, apparent policies, conditions, branches, and loops. Specific attention is placed on data entities. We assess their properties and methods to detect relationships across the entities in a given microservice codebase and extract the underlying data model.


Next, we continue with the \textit{construction} phase and convert the microservice-specific information into a graph format. This phase operates with components identified when traversing call-graphs. 
We further augment recognized components with additional information that might co-exist at the component definition level. For instance, REST controller endpoints might enforce access rights \cite{JSR:375,das2021automated}. 
Paying attention to components corresponds to the microservice development practice \cite{carnell2021spring}. In other words, the microservice will always process data and provide endpoints. Therefore, aggregating components and combining the call paths represents a graph, which we use as an intermediate representation of the processed microservice.

The \textit{manipulation} phase deals with the fusion of multiple microservice intermediate representations. As depicted in the previous discussion, we use two main ingredients: combining overlapping data entities and inter-microservice endpoint interactions. This is highlighted on \fig{merge}.
However, these strategies can be further extended, for instance, by information from build and deployment scripts. Moreover, the event-based approach with message brokers, such as Kafka or Messaging Queues, could be integrated here.

We begin the fusion by entity matching, specifically by looking for entities from distinct modules with a subset match of properties, data types, and possibly names. For this matching, we considered natural language processing strategies (Wu-Palmer algorithm \cite{han-etal-2013-umbc}). Then, combining all involved microservices, we derive the canonical data model (context map), and, through the matched entities, we promote data and control dependencies. 

The second ingredient considers inter-service interaction. First, we identify all endpoints, parameter types, and metadata, and then the remote procedure calls within the methods \cite{das2021automated}. Next, we match them, generate a complete system service overview, and augment the canonical model resulting from the previous strategy. 


Our manipulation phase results in a combined intermediate graph representation for the holistic system. This broadens the perspective for the consequent analysis with access to the canonical data model, inter-service dependencies, and the overall system service endpoints. It also maintains the specifics of each microservice, such as its bounded context with overlaps, technology information, and aggregate list of the technologies, broken up by layer, in the centralized perspective. Finally, since each microservice contains the build, deployment, and operation information, it allows the centralized perspective to render a graph of connected deployments.


The final \textit{analysis} phase of the SAR process is reasoning about the whole system. 
This article limits the discussion to our new overall architectural visualization process, described next.

To perform SAR for microservices, we implemented the Prophet tool\footnote{The code for the Prophet utility can be found at GitHub \url{https://github.com/cloudhubs/prophet-utils},\url{https://github.com/cloudhubs/prophet-utils-app},\url{https://github.com/cloudhubs/prophet}}. It follows the steps detailed in this section and performs static code analysis of Java-based source code. It recognizes component-based constructs behind Spring Boot and Enterprise Java. It utilizes the Java Parser library \footnote{\url{https://javaparser.org}} and a graph database (Neo4j)\footnote{\url{https://neo4j.com}} to store the microservice and system holistic results. 

The result is an intermediate graph representation of the system accessible through REST API. This representation can be used for system reasoning. In general, any kind of reasoning could operate with the intermediate system representation. For instance, we could detect design smells or security policy violations \cite{das2021automated,walker2020smells}. 
However, overall system visualization is best suited to facilitate reporting, facilitate navigation, and improve comprehension.

The microservice codebase has the most up-to-date system details. With an intermediate graph representation of the system received from static analysis applied across the decentralized codebases (i.e., through the continuous integration pipeline), any update can be reflected in a reconstructed system-centric view.



\section{Visualization of Microservices}
Many means can be used to articulate the reconstructed system architecture to stakeholders such as architects, developers, or DevOps. However, appropriate visualization can speed up comprehension of the reconstructed system and lead to expedited assessments of dependencies, bottlenecks, architectural smells \cite{walker2020smells, Taibi2018closer18} (i.e., poor design choices and anti-patterns), or consistency errors. 

This article considers two approaches: a conventional architectural visualization and a 3D visualization. We will assess each approach based on the following 
capabilities: \textit{Visualization ability}, \textit{Comprehensibility}, \textit{Navigation}, and \textit{Interaction~of~services}.

The SAR process may result in multiple architectural views. However, it is sufficient to limit our attention to a few views for a proof of concept. 
Thus, we pick the most beneficial views for microservices, those that support a system-centric perspective.

Mayer and Weinreich \cite{mayer_weinreich_dashboard} identified that supporting a view of service APIs and their interactions should be one of the most important goals that a tool designed for microservice analysis should achieve. When we consider Rademacher et al. \cite{10.1007/978-3-030-49418-6_21}, focusing on the service view is well justified. The service view defines the microservice's APIs and the inter-service calls between them. Furthermore, this view is also relevant for developers seeking to understand how the system~operates. 

Rademacher et al. \cite{10.1007/978-3-030-49418-6_21} also focused on the domain view. This view 
defines the domain model used by the microservice system, also known as the canonical model or context map. This view is necessary because microservices do not depend on a formal specification of a domain model, with each service instead of operating on its own bounded context, where it operates on the subset of entity attributes it needs \cite{msa}. 

Both service and domain views give a view of the system architecture as-is, showing the communication between the services and the state of the domain entities in use. These views can be used as documentation for developers and DevOps. Architects can compare the current architecture against the planned system architecture and detect deviations. They can also use it as a first warning to detect if the architecture has drifted from the original plan.



\subsection{On service and domain view information}


To extract necessary system information to construct the service view, we require two things: first, to detect the endpoints of each service, and second, to detect the calls made from one service to another using these endpoints. Software frameworks often provide utilities for quickly defining these endpoints in code, and this has been utilized by projects like Swagger\footnote{\url{https://swagger.io}} for automated endpoint documentation. After endpoints are identified, Prophet inspects the microservice Abstract Syntax Trees (ASTs) for remote method calls. These can be recognized through common constructs such as REST templates, etc. Once the list of endpoints and calls is collected for each service, Prophet matches the calls to system endpoints based on the relative endpoint URL, the HTTP method, and parameters. The result is a call-graph representing the system, showing how the services communicate among themselves.

To extract system information to determine the domain view, we need to identify data entities, their properties, and their relationships. 
Data entities use frameworks utilities and can be identified similarly to endpoints. Having entities identified we can consider their properties and relevant data types. Identified property data types can reveal relationships the entities have with each other. These relationships have three different components, which we extract using code analysis: the types involved in the relationship (i.e., the entities that are on either side of the relationship), the multiplicity of the relationship, and the directionality of the relationship. Identifying the types is done based on the type names of the entities' fields, the multiplicity can be determined by whether or not the field is a collection, and its directionality can be determined by whether or not there is a corresponding field in both of the entities involved or in only one entity.
Considering a single microservice codebase, we can derive a microservice bounded context. 

Using the bounded contexts for all microservices, a combined canonical model for the entire system can be generated by merging the bounded contexts. Since the services should be operating on some of the same entities, the entities in each microservice can be merged by detecting if they have the same or similar names. Different services may have different purposes for the entities they share and so may retain different fields from each other. Fields with the same or similar names and the same data type are merged into a single field in the merged entity, while non-matching fields from all the source entities can simply be appended to the merged entity. The result represents the scope of all entities used in the system.

\subsection{Considered system samples}

To demonstrate visualization approaches for this manuscript and on a large, realistic system, we adopted two test benches.
The Teacher Management System (TMS)\footnote{\url{https://github.com/cloudhubs/tms2020}} consists of three microservices, and the limited size allows us to embed complete SAR visualization examples in this article. For a demonstration of a complex case study, TrainTicket\footnote{\url{https://github.com/FudanSELab/train-ticket}}\cite{trainticket} is used (originating from the ICSE conference). The TrainTicket was designed to emulate a real-world microservice system consisting of 41 microservices and over 60,000 lines of code. It is written in Spring Boot, uses MongoDB as its database, and follows cloud-native practice with containers, routing, etc.


\subsection{Conventional architectural visualization and its properties} \label{conv}



\begin{figure*}[t]
	\centering
	\begin{minipage}{.95\columnwidth}
		\centering
		\includegraphics[width=0.9\textwidth]{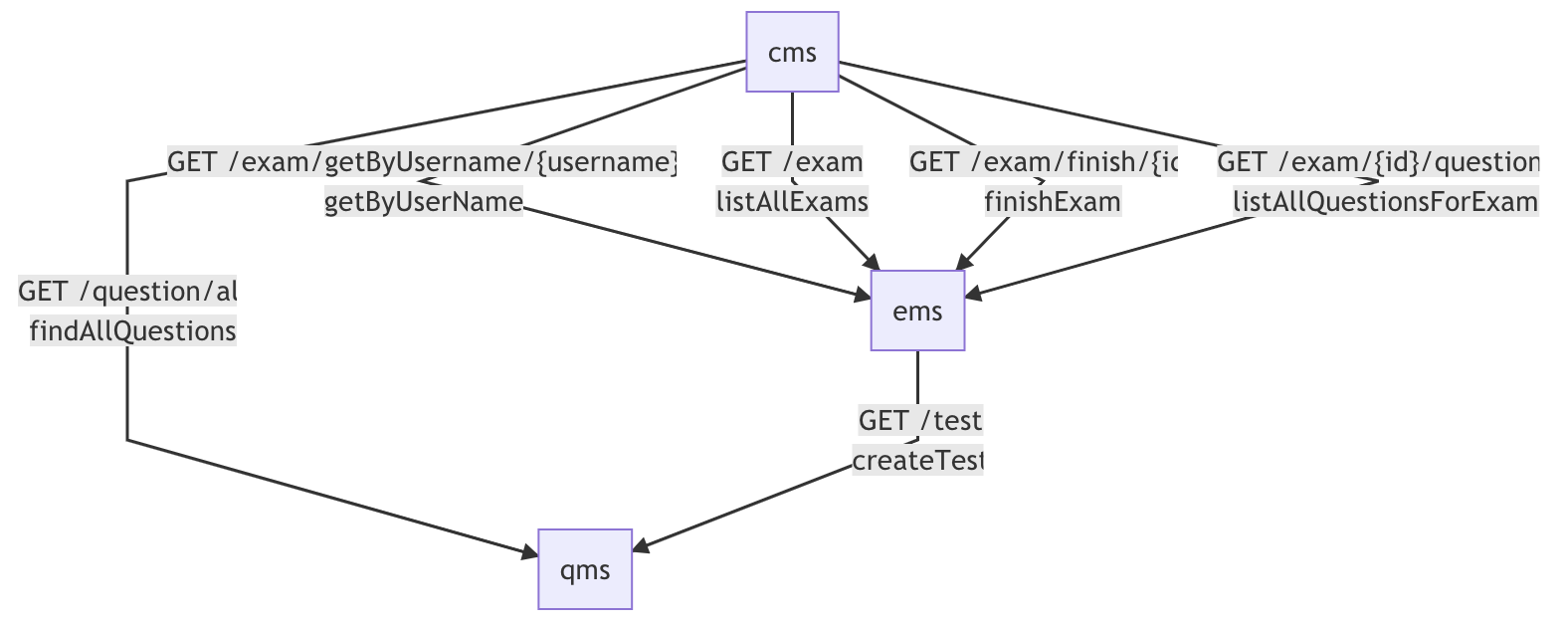}
		\caption{Sample service view extracted from a the TMS benchmark.}
		\label{fig:prophet}
	\end{minipage}%
		\begin{minipage}{.2\columnwidth}
		\end{minipage}
	\begin{minipage}{.95\columnwidth}
		\centering
		\includegraphics[width=\textwidth]{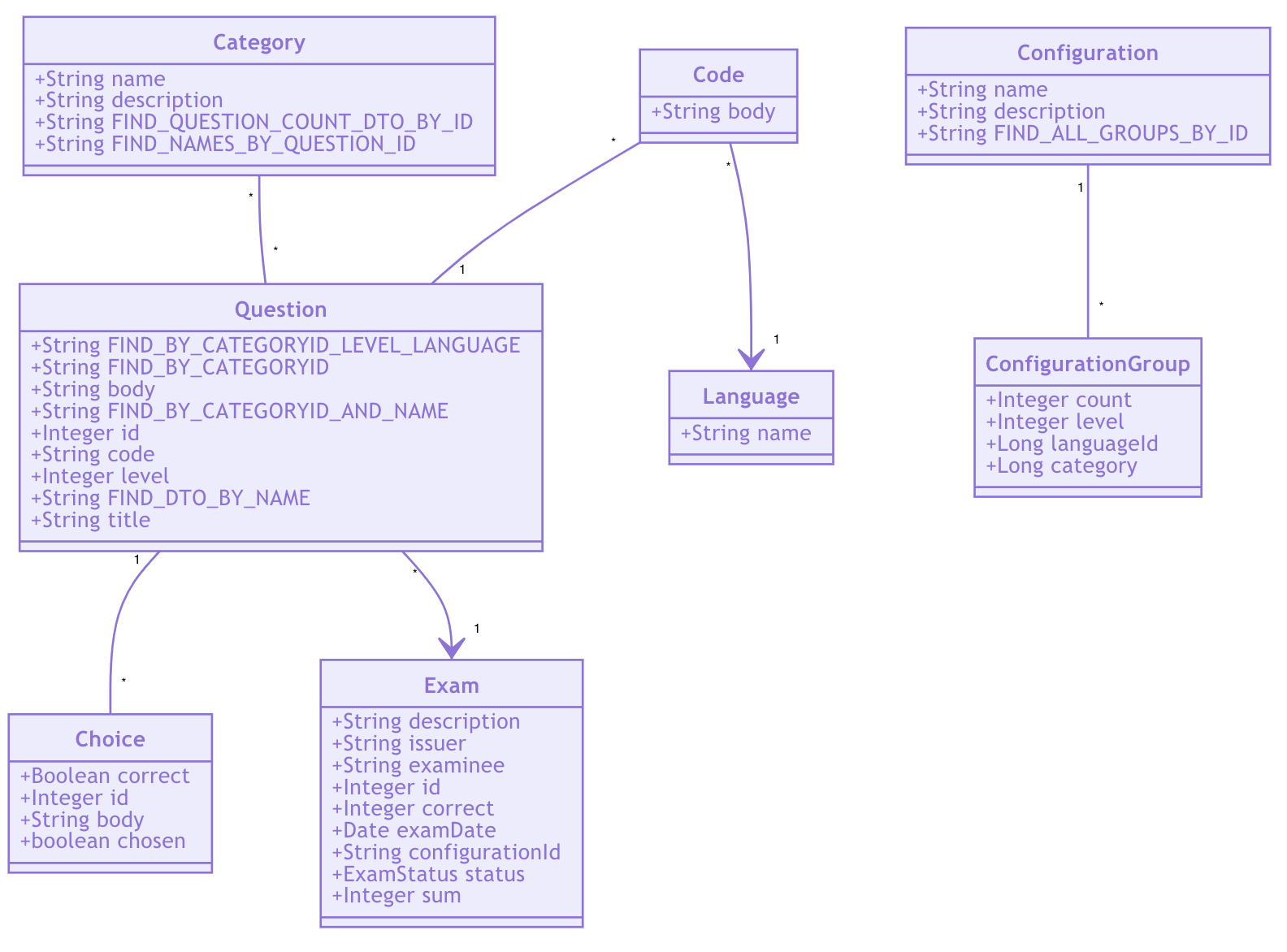}
		\vspace{-1em}
		\caption{Domain view derived from the TMS benchmark. These entities are aggregate definitions from partial entities in each microservice's bounded context.}
		\label{fig:prophet-domain}
	\end{minipage}
\end{figure*}




The conventional approach to visualizing service and domain views operates in two-dimensional space. The service view represents microservices as nodes and particular service calls as edges. An example output of the result of this analysis on the TMS testbench is shown in \fig{prophet}. 

The domain view has a perfect fit for the UML class diagram that represents the scope of all entities used in the system, as shown in \fig{prophet-domain} for the TMS system.


These results on the TMS system demonstrate a system-centric perspective extracted from the microservice codebase. 
Since this article focuses on visualization aspects, we next consider deficiencies and limits of obtained results. 

\begin{figure*}[ht]
\centering
\includegraphics[trim=0 75 35 100, clip,width=0.9\textwidth]{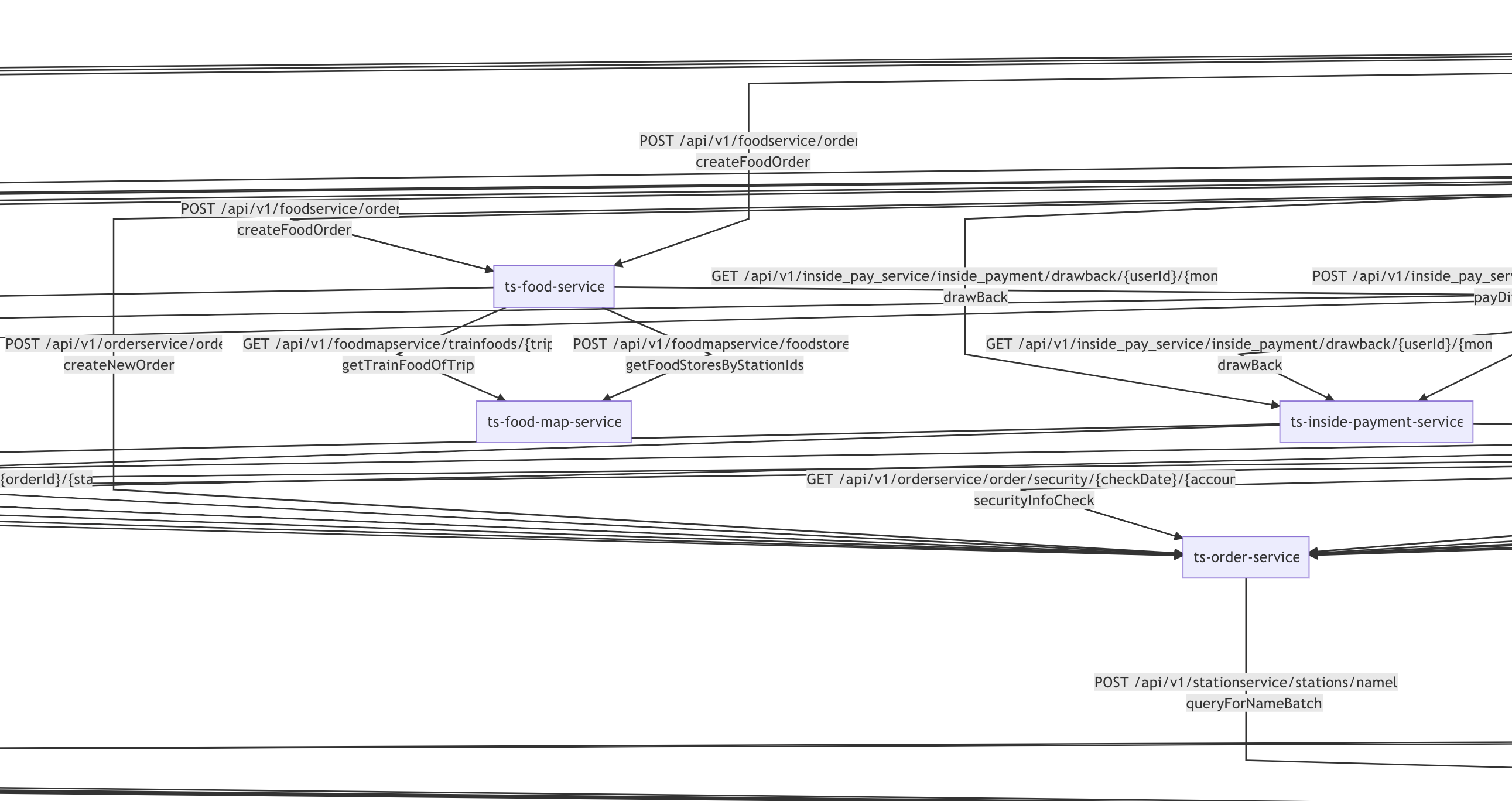}
\caption{The service view from a large microservice testbench TrainTicket \cite{trainticket}. Connections between services become difficult to decipher as the system size grows.}
\label{fig:prophet-cluttered}
 \vspace{-2em}
\end{figure*}

The biggest shortcoming of the conventional two-dimensional graph representation is its visualization ability; it quickly runs into scaling problems. We discovered that the visualization breaks down when analyzing systems larger than a few microservices. A two-dimensional space only has so much area available to display a graph, which fills up quickly and becomes unintelligible. This limitation is not surprising; as the number of services in a system increases, the potential number of connections between them increases at a much faster rate. There is only so much space in a two-dimensional layout to arrange these connections, and thus the visualization becomes cluttered and unwieldy. We discovered this problem when analyzing larger systems; \fig{prophet-cluttered} shows service view output on the TrainTicket testbench (41 microservices), which becomes difficult to understand.

A problem of visualization may seem like a minor one, but it directly affects the view's intended purpose as an artifact to help a stakeholder quickly understand how microservices interact with each other in a large system and to allow them to visually identify potential problems with the architecture or to identify drift from the originally-intended architecture. As the graphs become cluttered, this kind of quick visual analysis becomes less feasible, as it takes more time to understand what the graph is displaying. A visualization solution based on two-dimensional diagrams simply does not scale well with the number of microservices in a system.

The related problem is that of navigating the displayed graphs. While a small system can be displayed on a single page without much issue, the output requires users to navigate larger graphs using the mouse scroll wheel and does not provide for zooming in or out, nor any other method of viewing multiple levels of abstraction, an important feature of microservice architectural analysis as seen in, e.g., the hierarchical C4 model \cite{c4}. This limited method of navigation creates a problem since there is no way to step back and get a broad view of the system, nor can the user quickly drill into a specific region of the microservice mesh. It can take time and effort to find the area of interest in the displayed graph, and it may not be as insightful if developers cannot easily relate what they are looking at to the rest of the system. Again, this directly impedes the original goal of the quick and intuitive~analysis.

Another problem is that the information about each microservice's API is not easily accessible. The endpoints are only displayed on the edges that point to the node. The user must mentally reconstruct what the API looks like for a particular service by finding all of the incoming edges and identifying their labels. This is extra work for the user, which is also detrimental to the goal of quick visualization, and the difficulties with navigation as previously mentioned compound~the~task.

The final problem with the conventional visualization is its inability to display how the microservices interact with each other when servicing actual requests from users. Its visualization is completely static; the connections between services are there, but there is no information on how those connections are utilized. This also hinders the goal of providing at-a-glance visualization of a system; the static view of the connections provides only a partial picture.

To summarize, we identified these challenges:

\begin{enumerate}[leftmargin=.5cm]
    \item Visualization ability: the method needs to scale better with system size than a two-dimensional, UML-based solution.
    \item Comprehensibility: developers should be able to quickly comprehend the interaction of microservices in a system.
    \item Navigation: the visualization should be easily navigable and enable traversing multiple levels of abstraction.
    \item Interaction of services: a method is needed that can visualize how the services operate and interact with each other, beyond what is capable with UML-based diagrams (e.g., sequence diagrams).
\end{enumerate}

\subsection{A Microvision} \label{sec:Microvision}

The most applicable view for understanding the system-centric perspective and the system operation is the service view. 
With the 2D limitation in mind, we have adopted this view to explore the benefits of a three-dimensional visualization scheme. We utilize the AR medium, which is natively three-dimensional, and it lends itself to control schemes based on natural movement. We believe this combination holds potential for use with displaying and navigating complex systems such as microservices. The way we approach the visualization is by using a 3D graph operating in AR. Given we were able to automate the SAR process and reconstruct the service view in 2D, we use the same input for a 3D graph operating in AR.

The potential of using AR for software visualization has been recognized, and it has been used to visualize monolithic software systems in various ways as shown above in the Related Work section. However, it has not been previously applied to microservice systems. We aim to expand the existing 3D visualization techniques to a higher level of abstraction beyond a single piece of software to an entire distributed microservice system.




\subsubsection{Designing 3D Visualization}

For the {\it system-centric perspective}, the view needs to provide a {\it high-level system} visualization. In microservices, we intend to see their {\it interconnection}; however, the view cannot get cluttered as more services are introduced. Quickly understanding the high-level structure should be prioritized in all system-centric perspectives.  

In addition, the user should easily {\it interacts} with the view and {\it navigates} through the microservice system both at a high level and a lower level of detail centered around a few services. The high-level view should be easily understandable, and upon drilling down into a lower-level view, the user should be able to identify details about individual services and how they relate to their neighbors. 

In this case, the high-level view refers to the  {\it overall structure of the system}, and the low-level view refers to information about  {\it individual services} and their immediate neighbors. The less time it takes to go from high to low level of detail, the easier it is to understand the system and the roles the individual services play in it.


Given our ambitious AR microservice visualization plan, we have developed a Microvision proof-of-concept tool delivering the service view. Microvision consists of two broad components to achieve its functionality: the main {\it graph display, and the API viewer}. The following section describes the design of these components and details the rationale behind the components.

\begin{figure}[t]
\centering
\includegraphics[width=.47\textwidth]{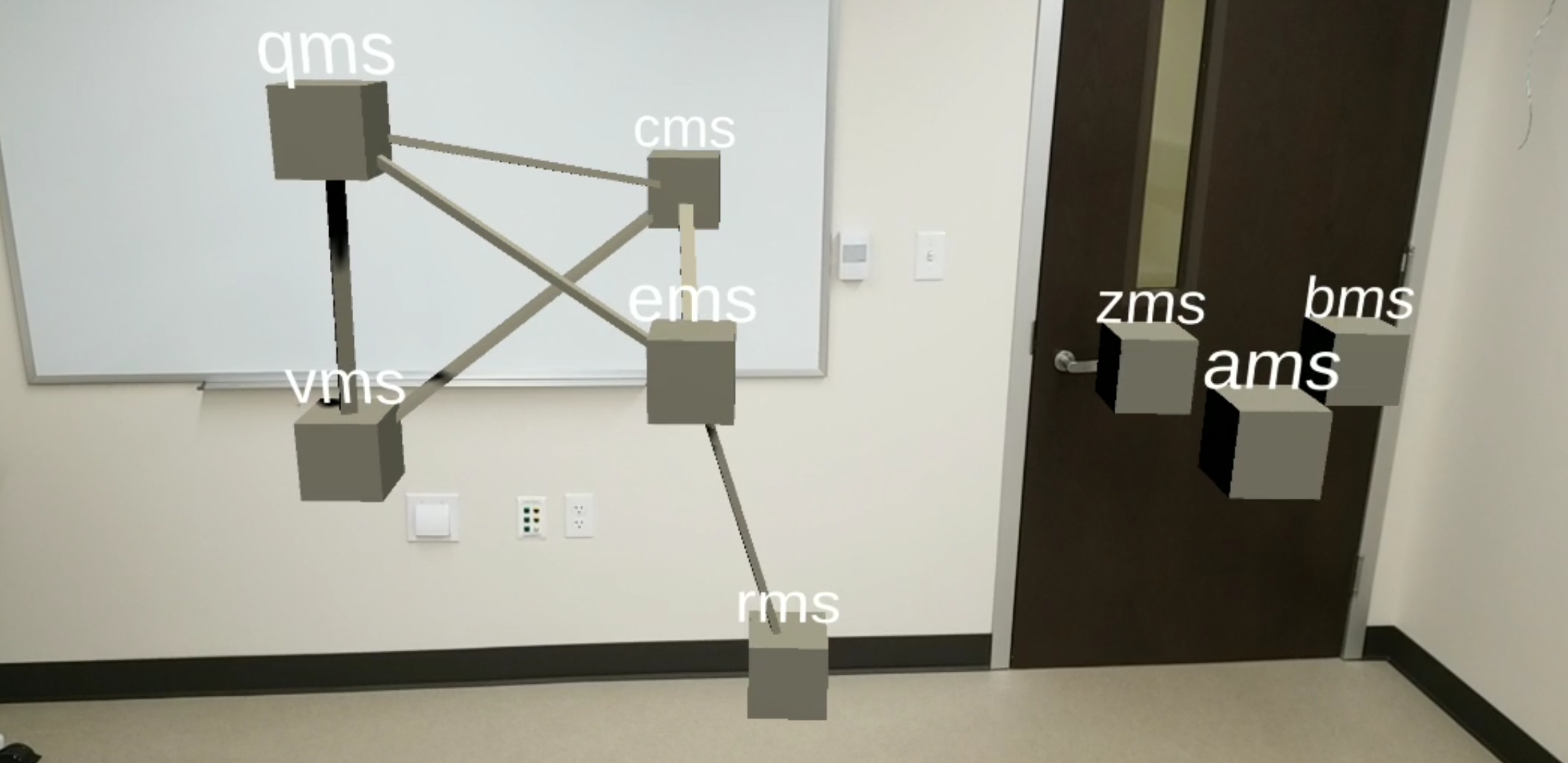}
\caption{Showing 3D graph of services and their connections.}
\label{fig:microvision-graph}
 \vspace{-2em}
\end{figure}

{\bf Graph display:} The overall display of the microservice system shows an abstract 3D graph view of the system projected in AR. The goal of the base graph is to give a quick view of the system, its services, and its connections. Each microservice is represented as a node, and an edge exists between nodes if a call exists between two microservices.
\fig{microvision-graph} shows our implementation example.
The nodes are distributed such that there is no overcrowding in one particular area of the graph. To help visualize how the connections work, a node can be selected to highlight it and its neighbors. In this case, the neighbors highlighted are only those that are called at some point from the selected node. This helps clarify the node's role among nearby nodes.

The rationale for choosing a graph display is simple. A graph is the most natural way of conceptualizing how microservices work, and two-dimensional graphs are consistently encountered in other microservice analysis tools.


The three-dimensional display addresses both of these drawbacks. First, adding the third dimension increases the available area to display the services and reduces the amount of overlap their connections have. This decluttering also makes it more straightforward to analyze the architecture and identify potential architectural problems or areas of drift from the original architecture. Second, displaying the graph in an AR environment allows us to implement a natural navigation scheme: the user simply moves their device through the graph. Since a 3D graph has an innate spatial logic to it, it is intuitive to navigate by natural motion, even for a large graph. This natural movement also allows for quickly switching between a broad overall view of the system or a closer look at a particular group of services simply by moving closer or farther.

The other alternative considered is to display the content of the system using a visual metaphor, such as ``software cities" or similar approaches encountered in prior work \cite{vr_software_cities, vr_evostreets, ar_software_islands}. We ultimately rejected this approach in favor of an abstract approach, since learning a new visual metaphor would take extra time and training, and the idea of how microservices connect is served well by an abstract graph representation. Furthermore, these visual metaphors usually restrict the space in which the services can be displayed; for example, the software city metaphor requires a two-dimensional layout of the system in question, with the vertical dimension being used to display information about the content of the software packages. For a large system of microservices, this space would be better utilized by simply displaying more services in a smaller area.

\begin{figure}[t]
\centering
\includegraphics[width=.47\textwidth]{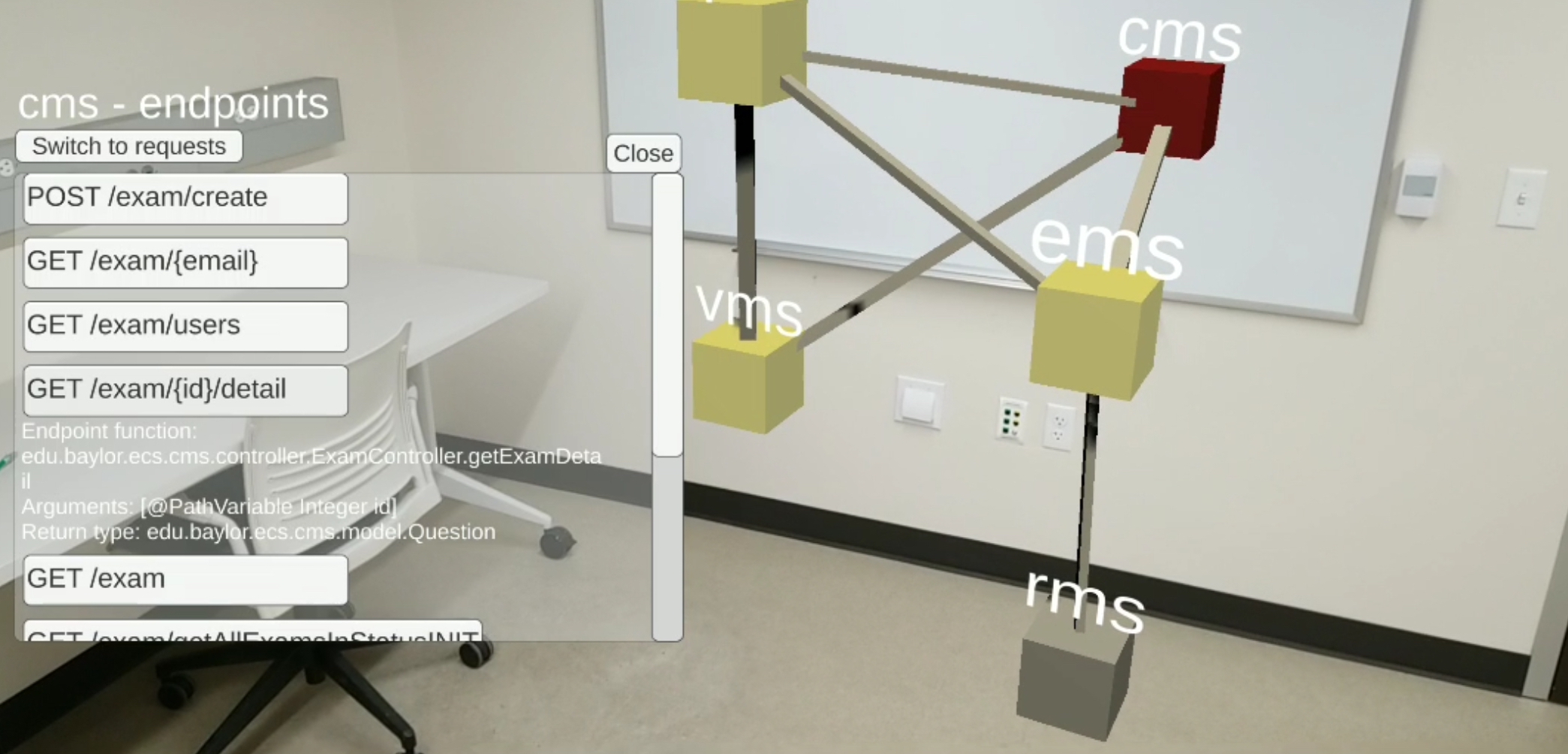}
\caption{The context menu shows a selected services API endpoints, in this case the ``cms" service highlighted in red.}
\label{fig:microvision-endpoints}
\vspace{-2em}
\end{figure}

{\bf API view:} The API view component is responsible for displaying the endpoints that make up the API of a selected microservice. The design goals for this component were to concisely display the relevant API for a particular microservice without cluttering the overall graph. For this component, a simple pop-up box that contains the list of endpoints was chosen. \fig{microvision-endpoints} shows our implementation example.

Upon selecting a microservice in the graph, the API view box will pop up and display the list of endpoints. Each endpoint is identified by the endpoint path and HTTP method. The endpoints can be expanded by tapping them, providing further information such as the actual method that implements the endpoint and its parameters and return type.

This component is intended to provide details for developers to drill down into. 
The graph itself provides both a broad overview and details of the services through a contextual window panel. We chose to show specifically the API because that is a microservice's best indicator of its role within the system. It is not by default shown within the graph itself, as that would add too much information that may not be immediately relevant to the user.

The primary alternative considered for this component is to annotate the node in the graph itself with the information. This way, the information could be positioned spatially in relation to the nodes it applies. This alternative was rejected because it would clutter the graph unnecessarily. The graph already contains the nodes, their connections, and the names of the microservices annotated onto the nodes. Adding extra text to the graph would make the existing elements more difficult to read and interpret. Furthermore, the convention of tapping an element to display an informational pop-up already exists, and since we highlight the selected node, we do not lose~any~clarity.

\subsubsection{Microvision Approach Summary}

Our proof-of-concept, Microvision\footnote{Its code is available at GitHub \url{https://github.com/cloudhubs/microvision}}, addresses the shortcomings of 2D visualization. We demonstrated 3D visualization based on AR to reduce the cluttered nature of the conventional visualization for microservices. In addition, we demonstrated navigation and control through the reconstructed microservice system architecture. Specifically, we addressed the challenges:

\begin{enumerate}
    \item Visualization ability: we have developed a 3D visualization that offers better scaling with the number of services than a 2D diagram.
    \item Comprehensibility: the 3D structure can be viewed for a high-level system overview.
    \item Navigation: the graph is displayed in AR and is easily traversed by natural movement. Multiple levels of abstraction are viewed naturally within the graph itself due to the 3D overview. 
    
\end{enumerate}

Still, the interaction perspective can be further developed with simulated endpoint interaction to better address dependencies and tracing.

\section{Small Evaluation Study} \label{sec:study}

A small-scale pilot study in terms of the number of participants and the number of microservices is included. It was conducted to evaluate the feasibility of the Microvision approach by answering practical developer questions and analyzing the architecture of a microservice system. We conducted the trial user study with graduate student volunteers performing various analysis tasks on a real-world microservice system using Microvision and giving feedback on their experience.

Our evaluation was conducted with a group of six graduate computer science students (five males, one female). All of the participants had experience with software development, and four participants had prior experience with microservices.

The participants were given a system consisting of 16 microservices to analyze. The system in question was a TrainTicket testbench \cite{trainticket} handling train ticket reservations. We prepared a set of nine evaluation tasks relating to this system for the participants to complete. Three questions were prepared relating to individual services and their connections in the system, and six were prepared relating to user requests and how the system handled them. The tasks required the participants to use Microvision to identify different aspects of the microservice APIs and their connections to each other. The tasks are given in Table \ref{tab:questions}.

\begin{table}[h]
\vspace{-.3em}
\begin{tabular}{|p{0.96\linewidth}|}
\hline
\textbf{General questions} \\ \hline
Which service has the most connections? \\ \hline
Of those connections, how many calls from that service to another? \\ \hline
How many services have only a single connection? \\ \hline
\textbf{Request \#1} \\ \hline
How many services are involved in this request? \\ \hline
What is the last call in the call chain? Include the service and endpoint. \\ \hline
Suppose the ts-ticketinfo-service changes the arguments required for its controller method, queryForStationId. Will this change affect this request? \\ \hline
\textbf{Request \#2} \\ \hline
What data type is the argument passed to the second endpoint? \\ \hline
What is the controller method that handles the initial request? \\ \hline
Suppose the ts-travel-plan service could be made to skip the route-plan-service and call the ts-travel-service directly. Would this change make the ts-route-plan-service obsolete in the system? \\ \hline
\end{tabular}
\caption{\label{tab:questions} Tasks completed by participants in the evaluation.}
\vspace{-1em}
\end{table}


We also prepared a 5-question satisfaction survey regarding participant experience with the application. The first three questions were given on a 5-point Likert scale and measured the participants' satisfaction with the use of Microvision on the specified tasks. The next two questions asked about the participants' perceived usefulness of Microvision's features and asked for suggestions for new features. The feedback survey and the participant responses are given in \tab{feedback}.


The evaluation was split into three segments. In the first segment, the participant was briefed on the features of Microvision and given directions on how to operate it. This segment lasted a maximum of ten minutes. In the second segment, the participants were given 15 minutes to complete the evaluation tasks, beginning with the three tasks relating to individual services, followed by the six tasks relating to user requests. The tasks were given one at a time, with immediate feedback as to whether the participant answered correctly or not. In the final segment, the participants were given the feedback survey, which they could complete in any amount of time they chose.

\subsubsection{Evaluation Results}

\begin{table}[h]
\vspace{-.3em}
\begin{tabular}{|p{0.5\linewidth}|p{0.38\linewidth}|}
\hline
\multicolumn{2}{|l|}{\textbf{Satisfaction feedback (5-point Likert scale)}} \\ \hline
Was Microvision helpful to you when completing the tasks? & 4 strongly agree, 2 agree \\ \hline
Was Microvision intuitive to use? & 4 strongly agree, 2 agree \\ \hline
Given the option, would you use Microvision again? & 4 strongly agree, 2 agree \\ \hline
\multicolumn{2}{|l|}{\textbf{General feedback}} \\ \hline
Which features, if any, did you find helpful or useful when using Microvision? & 3D graph visualization: 6/6; API viewer: 5/6 \\ \hline 
Do you have any other comments or suggestions regarding Microvision? & Graph scaling and individual node movement suggested \\ \hline
\end{tabular}
\caption{\label{tab:feedback} Feedback questions posed to participants.}
\vspace{-1em}
\end{table}

All six of the participants completed the evaluation tasks with 100\% accuracy within the allotted time frame. Furthermore, all the participants either agreed or strongly agreed to the first three qualitative feedback questions. Regarding the features, all participants said the 3D graph visualization was useful, and five out of the six participants said the API viewer was useful. These results show that Microvision is a promising direction to further research for microservice system analysis. 






    

\section{Conclusions} \label{sec:conclusion}

This research was motivated by recurrent microservice system challenges regarding missing system-centric views. It analyzed cloud-native systems using static analysis to demonstrate it is feasible to derive a system-centric perspective of these systems. It also elaborated on alternative visualization directions using three-dimensional space to render the system's service view in AR. We implemented a Microvision tool that uses the intermediate representation of the system built by the static analysis tool Prophet, which is capable of multi-codebase analysis for microservices. We assessed the approach on two system testbench with a short evaluation study to better understand the practical implications and potential impacts of such a visualization. In future work, we plan to perform a large user study for which we have already obtained IRB approval. The SAR process will also be generalized to a platform-agnostic approach as initiated in \cite{schiewe2022}. We will also assess more architectural views and alternative 3D models.



\section*{Acknowledgment}
This material is based upon work supported by the National Science Foundation under Grant No. 1854049, grant from Red Hat Research, and Ulla Tuominen (Shapit).

\bibliographystyle{IEEEtran}
\bibliography{access}

\end{document}